# Doping Controlled Superconductor-Insulator Transition in $Bi_2Sr_{2-x}La_xCaCu_2O_{8+\delta}$

Seongshik Oh[*], Trevis A. Crane, D. J. Van Harlingen, and J. N. Eckstein

*Department of Physics, University of Illinois, Urbana, Illinois 61801, USA*

We show that the doping-controlled superconductor-insulator transition (SIT) in a high critical temperature cuprate system ($Bi_2Sr_{2-x}La_xCaCu_2O_{8+\delta}$) exhibits a fundamentally different behavior than is expected from conventional SIT. At the critical doping, the sheet resistance seems to diverge in the zero temperature limit. Above the critical doping, the transport is universally scaled by a two-component conductance model. Below, it continuously evolves from weakly to strongly insulating behavior. The two-component conductance model suggests that a collective electronic phase separation mechanism may be responsible for this unconventional SIT behavior.

---

[*] soh4@uiuc.edu, Present address: National Institute of Standards and Technology, Boulder CO 80305.

Since the late 1980s, the superconductor-insulator transition has been intensively studied both experimentally [1-4] and theoretically [5-9] in homogeneously disordered two-dimensional (2D) thin films. These studies suggest that there exists a defect-scattering dominated metallic (that is, resistance being temperature-independent) state with a finite zero-temperature critical sheet resistance at the transition from superconducting to insulating behavior. Samples with sheet resistance less than this critical value are superconducting as T→0, otherwise they are insulating.

High-$T_c$ cuprates (HTCs) are a quasi-2D system, and magnetic field [10,11] or doping control [12-14] can continuously transform superconducting cuprates into insulators as in the conventional SIT. In the case of doping-controlled SIT, there exists a common feature for all cuprates in the temperature dependence of the in-plane resistance. For doping levels much higher than the critical doping, the resistance monotonically decreases as a function of decreasing temperature and drops to zero below a superconducting transition temperature. For doping levels below the critical doping, the resistance first decreases as a function of decreasing temperature at high temperatures, reaches a minimum at an intermediate temperature and then it starts to increase as the temperature is further reduced. However, for a small range of doping levels just above the critical doping, the temperature dependence of the resistance is composed of three regions [10-13]. As a function of decreasing temperature, the resistance first decreases at high temperatures, reaches a minimum at an intermediate temperature, starts to increase at lower temperatures until it reaches a local maximum, and then it drops to zero resistance below the superconducting transition temperature. Although such a re-entrant behavior is universally observed in almost all underdoped cuprates [10-13], its origin or significance has not attracted much attention so far.

In this Letter, however, we report the results of a doping-controlled SIT study with an unprecedented level of detail, and we find that important physics has been hidden in this phenomenon. First of all, the sheet resistance tends to diverge as T → 0 at the critical doping, contrary to the common notion of finite critical sheet resistance not only in conventional disordered metals [1-4] but also in cuprates [12,13]. Above the critical doping, the re-entrant behavior is well described by a two-component scaling function composed of a weakly insulating phase and a superconductive fluctuating phase. This model is consistent with the widely predicted and observed electronic phase separation mechanism driven by carrier-carrier correlations [15-19]. We attribute the observed divergence of the critical sheet resistance to this phase separation phenomenon. Below the critical doping, the transport continuously evolves into two-dimensional variable range hopping.

The samples used for this study are $Bi_2Sr_{2-x}La_xCaCu_2O_{8+\delta}$ (Bi2212:La=x) thin films grown by ozone assisted molecular beam epitaxy [20]. As the doping level of the samples is decreased, the temperature dependence of the sheet resistance, $R_S$, defined per $CuO_2$ bilayer, transforms from superconducting to insulating behavior as shown in Fig. 1. From Fig. 1, one may be tempted to estimate a zero-temperature critical sheet resistance ($R_{S0}$), close to the pair quantum resistance $\sim h/4e^2$, separating superconducting from insulating samples. However, in-depth measurements below show this not to be the case.

For a more detailed doping-controlled SIT study, we have grown a 40 molecular layers of Bi2212:La=0.44 film and varied its doping level in the vicinity of the SIT in small steps (smaller than 0.001 holes per Cu in doping) using precise and reversible oxygen control [20]. Each curve in Fig. 2 is obtained after a few minutes of either vacuum (~$10^{-8}$ Torr) or ozone (~$10^{-6}$ Torr) annealing at ~400 ºC. Fig. 2 shows sheet resistance vs. temperature ($T$) curves very near the SIT. Over all, there exist two temperature regions in Fig. 2(a). Each curve reaches a local minimum at a characteristic temperature $T_{min}$ near 80 K. Above $T_{min}$, all the curves exhibit metallic behavior ($dR/dT > 0$), and doping simply provides a temperature-independent shift with respect to one another. However, below $T_{min}$ all the curves switch to an insulating behavior ($dR/dT < 0$), diverging away from each other as temperature decreases.

When the temperature is further reduced, the temperature dependence of the resistance either switches to a metallic behavior ($dR/dT > 0$) when the doping exceeds a critical value or remains insulating. This is more easily seen in Fig. 2(b) where we plot resistance versus log$T$. We interpret the change to metallic behavior as the onset of superconducting fluctuations and expect these samples will eventually become superconducting at low temperatures since they are obtained by incrementally reducing the doping level of a sample which shows a superconducting transition above 4.2 K. The samples which exhibit superconducting fluctuations show a resistance peak $R_p$ at a temperature $T_p$ below which the resistance drops. Conversely, samples which do not show a peak in resistance are on the insulating side of the SIT.

In conventional superconductors, the temperature dependence of resistance just above the superconducting transition temperature is generally flat due to temperature-independent defect scattering [1,4]. However, emergence of a superconducting transition from an insulating temperature dependence, which we call re-entrant behavior, has been observed in granular superconductors, where superconducting islands are embedded in an insulating background [21-23]. In such a case, even when the high temperature dependence shows insulating behavior due to Coulomb charging energy, if the intergrain tunnelling energy and the Josephson coupling energy overcome the Coulomb energy at reduced temperatures, a superconducting transition can occur from an insulating normal phase. We propose that the insulating normal-state temperature dependence, which we see here even in superconducting samples, is of a similar origin. Although HTCs are structurally homogeneous, theories have predicted that hole carriers may segregate into hole-rich and hole-poor islands at low doping due to strong carrier-carrier correlation effect [24-27], and scanning-tunnelling experiments have verified that such granular electronic states prevail in underdoped HTCs [18,19].

According to this phase separation scenario, the re-entrant behavior observed for doping levels above the critical value can be qualitatively understood. At temperatures above $T_{min}$, strong thermal fluctuations wash out any phase separation mechanism, leaving only metallic behavior. Below $T_{min}$, however, phase separation occurs such that hole-rich (presumably superconducting) islands are formed inside a hole-poor (presumably insulating) background. In this case, the resistance is dominated by the insulating background until superconducting islands start to couple with each other at lower temperatures, ultimately giving rise to a global superconducting transition. Although such re-entrant behavior is universally observed in almost all underdoped HTC systems [10-13,28], its origin has not been well-known. Below, using a two-component conductance model, we present a quantitative analysis supporting the phase-separation as the main mechanism behind this re-entrant behavior.

A key finding of our work is that the bottom 8 curves (curves 6 through 13) from Fig. 2(b), which show resistance peaks above 4.2 K, can be rescaled by dividing the temperature by $T_p$ and the resistance by $R_p$. When this is done, the low temperature parts of the curves overlap nicely as shown in Fig. 3(a). This suggests that there exists a single scaling function $r = r(t)$, where $r \equiv R/R_p$ and $t \equiv T/T_p$, which describes the transition from insulating to superconducting temperature dependence for all of the superconducting curves. One can find a well-fitting scaling function, $r(t)$, using a simple two-component conductance model, where the total conductance, $g(t) = r(t)^{-1}$, is due to two contributions and is given by $g(t) = At^q + B\dfrac{t_c}{t - t_c}$ with $A = 0.881$, $q = 0.145$, $B = 1.72$ and $t_c\, (\equiv T_c/T_p) =$ 0.065 found as the parameters leading to the best fit of the data. Here $T_c$ represents the superconducting critical temperature, below which the resistance of the sample turns zero. Since $g(1) = 1$ and $g'(1) = 0$ by the definition of $T_p$, among the four parameters only two (for example, $q$ and $t_c$) are independent, and the other two can be obtained from $A = (1 + q(1 - t_c))^{-1}$ and $B = Aq(1 - t_c)^2$. The two independent model parameters, $q$ and $t_c$, determine the relative insulating and superconductive-fluctuating transport contributions respectively, as described below.

Both terms of this two-component conductance formula correspond to physically identifiable conductance channels and this is consistent with the above-mentioned phase separation mechanism. The first term represents the conductance of 2D weakly-insulating normal carriers from the hole-poor background, while the second is due to the conductance of fluctuating superconducting pairs from the hole-rich islands, similar in form to the 2D superconducting fluctuations treated by Aslamasov and Larkin (AL) [29]. In AL theory, the fluctuation contribution, however, is not referenced to any material property and is scaled only by $4e^2/h$. In our case, since $1/R = g(t)/R_p$, the contribution of this term to the overall conductance is scaled by $1/R_p$ which is strongly dependent on doping.

The two-component scaling function determined from curves 6-13 in Fig. 3(a) is plotted in Fig. 3(b) together with curves 3-13. Curves 4 and 5 can be nicely fit to the scaling function with $T_p$ values less than 4.2 K and corresponding $R_p$ values. Curves 1 and 2 cannot be fit to the scaling function for any choice of $T_p$ and $R_p$ values. Curve 3 can be scaled with a range of $R_p$ and $T_p$ pairs in which $T_p$ can take any values between 0 and ~0.5 K. In order to show this, we have selected three representative values, 0.1 K, 0.01 K and 0.001 K for $T_p$ and presented the corresponding fitting curves in Fig. 3(b). All three curves are equally well fit to the scaling function even if their $T_p$ values are orders of magnitude different with one another. This means that the doping level, $d$, in curve 3 is very near the $T_c = 0$ transition value, but it is impossible to know whether it is exactly zero or just very nearly zero.

The temperature dependence corresponding to the critical doping, $d_c$, can be obtained by taking the limit of the scaling function as $T_c \propto T_P \rightarrow 0$, which corresponds to $R = 1.14 R_p (T_p/T)^{0.145}$. Once the relationship between $R_p$ and $T_p$ is known, temperature dependence at critical doping can be determined. Our two-component conductance model does not give any *a priori* information about how $T_p$ and $R_p$ are related, and they should be obtained from $R$ vs $T$ curves. Fig. 3(c) shows the result of this, and we find $R_p \approx 1.72 \times T_P^{-0.149}$ as the best fit for the 8 data points (curves 6-13) of $R_p$ vs $T_p$. Values of $R_p$ and $T_p$ for the other two (curves 4 and 5) with $T_p < 4.2$ K are obtained by fitting the corresponding $R$ vs $T$ curves to the scaling function and are plotted for comparison as well. Using this relationship, the temperature dependence at critical doping reduces to $R = 1.96 \times T^{-0.145} T_P^{\alpha}$, where $\alpha$ = -0.004.

The exact value of $\alpha$ is significant because $\alpha = 0$ implies that there exists an asymptotic critical-doping temperature dependence to which $R(T)$ tends as the doping approaches $d_c$ from the superconducting side. Since the error in determining the exponent is close to 0.01, we believe the obtained value is effectively zero. If $\alpha$ were not zero, then as $T_c \rightarrow 0$, $R$ would tend either toward zero or infinity for all temperatures depending on the sign of $\alpha$, either of which is unphysical. Since we find $\alpha \sim 0$, we believe our scaling function and empirical relationship between $R_p$ and $T_p$ provide a well-behaved limiting form as $T_c$ goes to zero.

This result is interesting because it implies that, for the doping controlled SIT, the critical temperature dependence is insulating ($dR/dT < 0$), and a zero-temperature critical sheet resistance does not exist. This is in direct contrast to the conventional picture of the SIT, according to which a temperature-independent and finite sheet resistance exists at a critical value of the control parameter [1-4,12,13]. With respect to this interpretation, one may doubt if the analysis obtained from data above 4.2 K can be applied all the way down to the zero temperature. The best way to resolve this issue would be to extend the investigation down to much lower temperatures. However, we believe that the above conclusion would remain unchanged, because there is no indication in the transport data of any deviation from the two component scaling form in the temperature range we have studied.

Making use of the normal state conductance, $G$, which is monotonically related to the doping level as shown in Fig. 1, it is possible to estimate the doping level, $d$, of each curve, since within a reasonable approximation $G$ must be proportional to $d$. Assuming that $d$ of curve 3 is the critical value, $d_c$ and taking 0.05 hole/Cu, the widely-accepted value for the critical doping [30], as the critical value, we have converted $G$ at 150 K into $d$ in Fig. 3(d) and presented $T_p$ as a function of the doping level. Since $T_c$ ($\propto T_p$) is the energy scale that governs transport on the superconducting side and renormalization scaling predicts $T_c \propto |d-d_c|^{\nu z}$ [6], a power law relationship is expected between $T_p$ and doping. From Fig. 3(d), the estimated critical exponent is $\nu z = 1.8 \pm 0.3$.

Transport on the insulating side of the SIT is shown in Fig. 4, where curves 1 and 2 from Fig. 2 are reproduced. We also measured the temperature dependence of the sheet resistance for a barely insulating sample between 2 and 0.2 K; this is labelled curve *a*. For curves *a* and *b*, the resistance is well described by 2D variable-range hopping (VRH), where $R(T) = R_0 \exp(T_0/T)^{1/3}$ [31,32]. Here $T_0 = 10/k_B \xi^2 N(E_F)$ where $N(E_F)$ is the density of states at the Fermi level, and $\xi$ is the characteristic size of the localized states between which hopping occurs. In our data $T_0$ tends to zero as the doping level approaches the critical doping from the insulating side. Curve 2 is better fit by a power law within the measured temperature range. Curve 1 tends to deviate from a high-temperature power-law toward a low-temperature 2D-VRH dependence. Even curve 2 may take a 2D-VRH form at very low temperatures. This observation suggests the existence of a crossover temperature, $T_i$, at which a weakly insulating (power law dependent) high temperature behavior transforms into a strongly (quasi-exponential such as 2D-VRH) insulating low temperature one. In the other extreme, when $T_0$ gets very large and $\xi$ very small, the transport starts to diverge faster than is expected by VRH, which is the case for curves *c*, *d* and *e*.

All the above observations can be summarized into a ($T$, $d$) phase diagram shown in Fig. 5. It is composed of metallic, insulating, superconductive fluctuating (i.e. local superconducting), and global superconducting regions divided by three crossover temperatures, $T_{min}$, $T_p$ and $T_c$. The insulating region can again be divided into weakly and strongly insulating regions by $T_i$. Fig. 5 presents a detailed view right around the critical doping, which has not been well treated in the conventional HTC phase diagrams. According to the above analysis, this phase diagram is interpreted as follows. When the doping level gets lower than ~0.058 (hole/Cu), the high temperature homogeneous metallic phase starts to segregate into superconductive fluctuating islands and the sea of insulating background at temperatures below $T_{min}$. The resistance is first dominated by the weakly-insulating background. However, as the temperature is further reduced, the fast-increasing conductivity of the superconductive fluctuating islands overturns the insulating temperature dependence into metallic ($dR/dT > 0$) behavior below $T_p$. At even lower temperatures (below $T_c$), when Josephson coupling is formed among the superconductive fluctuating islands, a global superconducting state emerges. Along the doping axis, as the doping level is reduced from ~0.058 (hole/Cu), the relative portion of the superconductive islands gets smaller and the insulating temperature dependence survives down to lower temperatures. Eventually when the doping level reaches the critical doping, 0.050 (hole/Cu), the weakly-insulating behavior survives all the way down to the zero temperature and this results in absence of the zero temperature critical sheet resistance, which, however, needs to be verified by further studies at much lower temperatures, considering that our measurement was obtained only above 4 K.

The authors thank Philip W. Phillips, Anthony J. Leggett, and Michael B. Weissman for useful discussions. We also acknowledge the US Office of Naval Research for supporting under grant N00014-00-1-0840 and the use of the Frederick Seitz MRL-CMM at UIUC through U.S. DOE, Div. of Mat. Sci. award #DEFG02-91ER45439.

**Figure Captions**

Fig. 1 (color online). Doping controlled SIT over a wide doping range. The solid arrow on the vertical axis indicates the pair quantum resistance, $h/4e^2$.

Fig. 2 (color online). Detailed view of doping-controlled SIT over a very small range of doping levels. (a) and (b) are the same data plotted on different horizontal scales. Note that each curve is indexed by a number 1 through 13 in the order of increasing doping.

Fig. 3 (color online). Scaling of superconducting curves. Each index represents the corresponding curve from Fig. 2(b). (a) Rescaled plot of curves 6-13 from Fig. 2(b). (b) The two-component scaling function plotted together with curves 3 -13. Curves 6-13 are used to obtain the scaling function. Curves 4-5 are fit to the obtained scaling function with $T_p$ and $R_p$ as the fitting parameters. As for curve 3, any choice of $T_p$ below ~0.5 K gives reasonably good fitting; three curves with $T_p$ = 0.1 K, 0.01 K, and 0.001 K are shown from left to right as examples. (c) $R_p$ vs $T_p$ for the 10 superconducting curves (4-13). The solid line represents the best fitting for the 8 measured data points (curves 6-13), which corresponds to $R_p = 1.72 T_p^{-0.149}$. (d) $T_p$ vs doping. Conductance at 150 K is also shown on the top axis. The solid line, $T_p = 4.02 \cdot 10^5 \cdot (d - d_c)^{1.8}$, is the best fitting for $d > d_c$.

Fig. 4 (color online). $R_S$ vs $1/T^{1/3}$ for insulating doping levels. Linearity in this plot implies 2D-VRH. Curves 1 and 2 from Fig. 2 are reproduced for comparison. Inset shows how curve *a* compares with other curves from Fig. 2.

Fig. 5 (color online). $(T, d)$ phase diagram near the critical doping. $T_p$ and $T_{min}$ are determined by the resistance data, and $T_c$ is evaluated as $0.065 \cdot T_p$ according to the scaling analysis. Fitting curves for each of these crossover temperatures are plotted as well. Unlike other crossover temperatures, $T_i$ is only marginally defined in our experiment.

Fig. 1

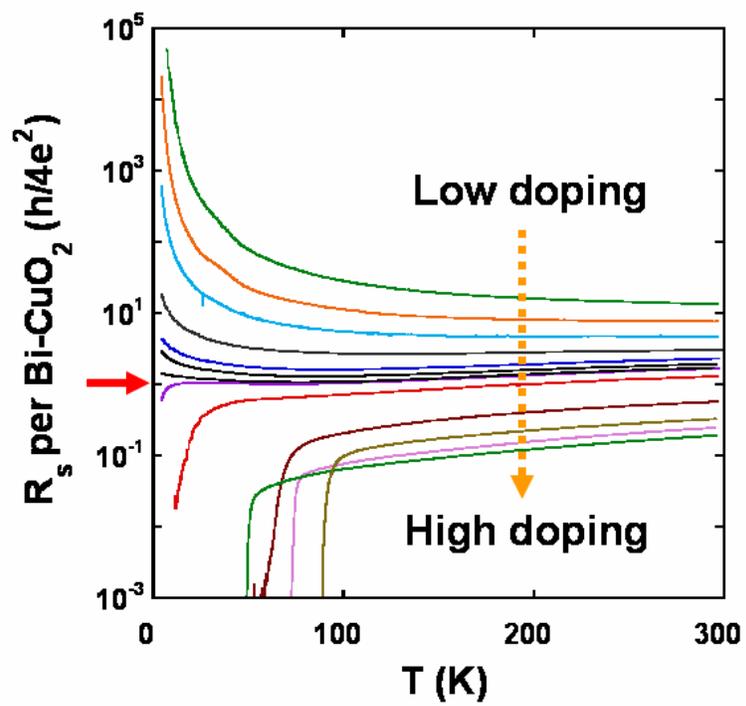

Fig. 2

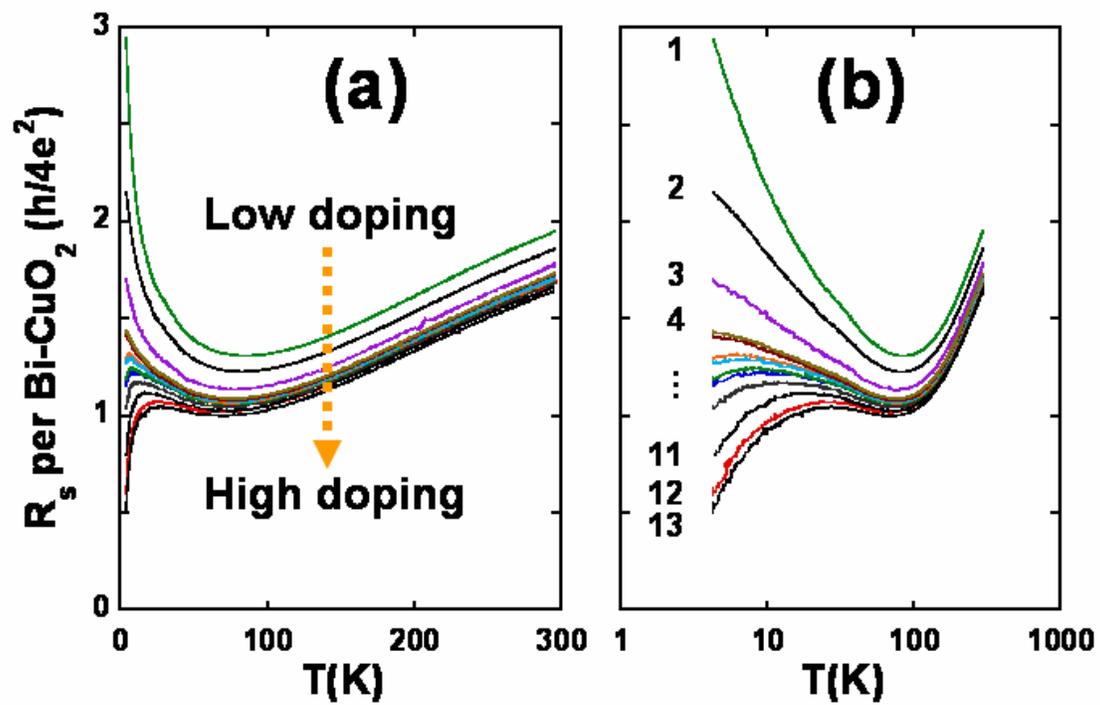

Fig. 3

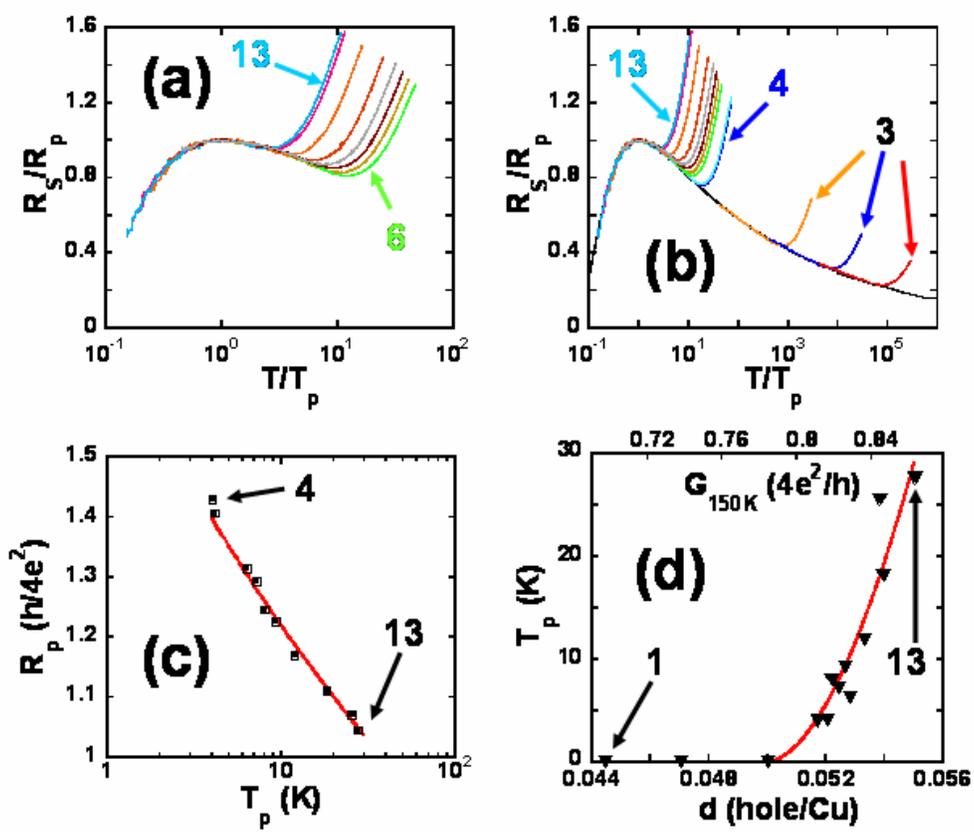

Fig. 4

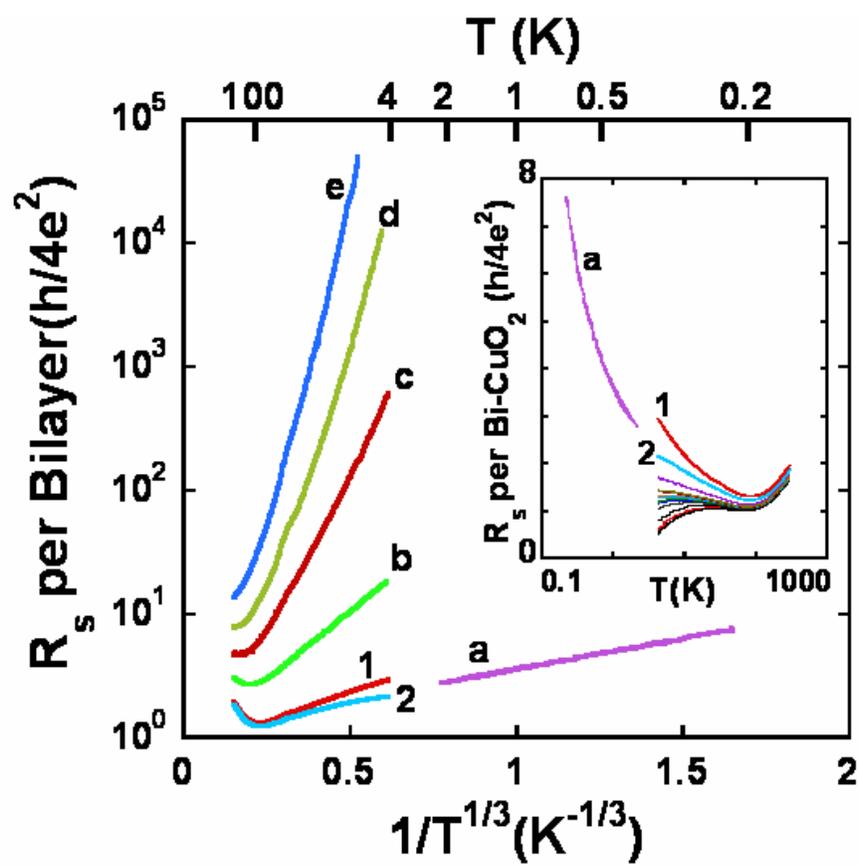

Fig. 5

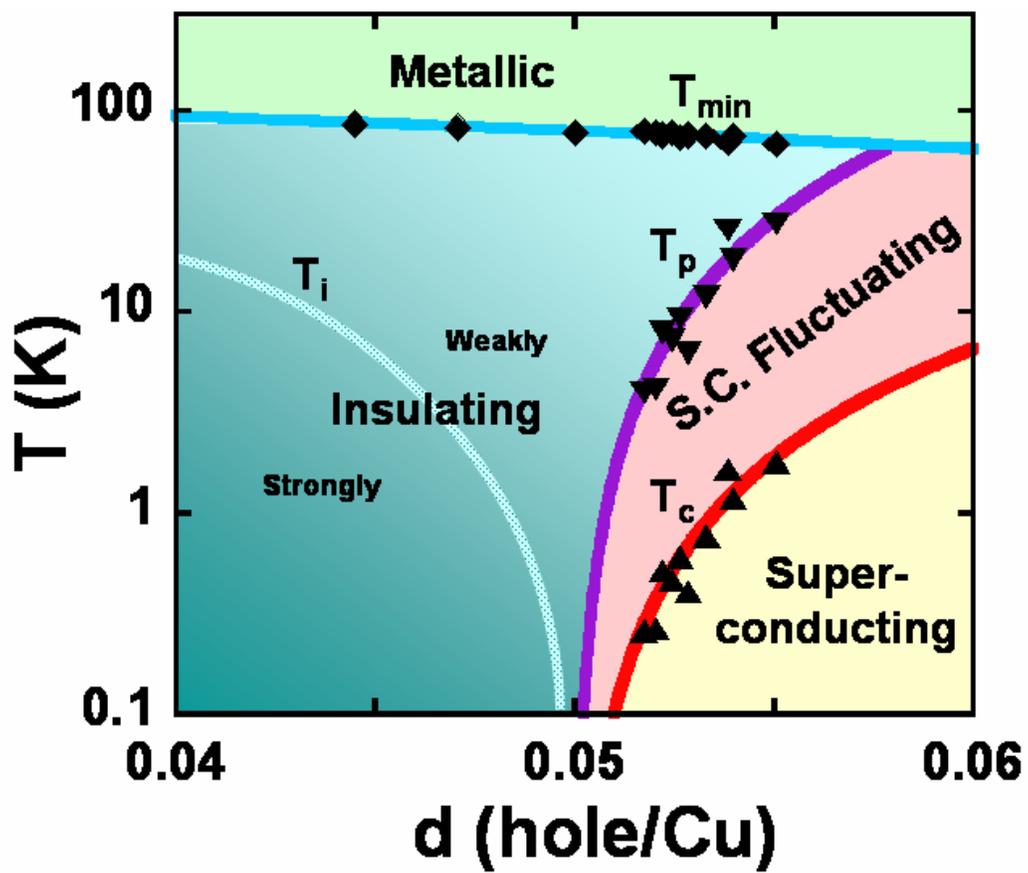